\newcommand{\beq}{\begin{equation}}
\newcommand{\eeq}{\end{equation}}
\newcommand{\be}{\begin{equation}}
\newcommand{\ee}{\end{equation}}
\newcommand{\bea}{\begin{eqnarray}}
\newcommand{\eea}{\end{eqnarray}}
\newcommand{\ena}{\end{eqnarray}}
\newcommand{\bear}{\begin{eqnarray}}
\newcommand{\ear}{\end{eqnarray}\noindent}

\newcommand{\bt}[1]{{\bar t}}

\documentclass[11pt]{article}
\input epsf
\usepackage{amsmath,amsfonts,epsfig,color,latexsym}
\usepackage{amssymb}
\csname @addtoreset\endcsname{equation}{section}
\begin{document}
\title{ Generation and Evolution of Spin Entanglement in NRQED}         % Enter your title between curly braces
\author{Ru-Fen Liu\footnote{fmliu@phys.ncku.edu.tw} {} and Chia-Chu Chen\footnote{chiachu@phys.ncku.edu.tw}}   % Enter your name between curly braces
\date{\textit{National Cheng-Kung University, Physics Department\\ Center for Quantum Information Science\\ 1 University Road, Tainan, 70101, Taiwan, R.O.C.}}               % Enter your date or \today between curly braces
\maketitle \normalsize

%\vskip1cm {\bf\Large \today \hskip5mm  File name apss$-$new2.tex} \vskip1cm

{\bf ABSTRACT}\\[.0015in]

\indent Analysis on the generation of spin entanglement from
non-relativistic QED is presented. The results of entanglement are
obtained with relativistic correction to the leading order of
$(\frac{v}{c})^2$. It is shown that to this order the degree of
entanglement of a singlet state does not change under time
evolution whereas the triplet state can change.
\newline \indent

\ \ \ \ \ \
%\setcounter{footnote}{0}

%\section{Introduction}
\ \ \ \ The generation of entanglement states is one of the
important problems of quantum information science. This is due to
the fact that entanglement is the essential source of quantum
computing and quantum information processing. An interesting
example of entanglement is the famous Einstein-Podolsky-Rosen
(EPR) pair\cite{Ein} which has been suggested to provide secure
quantum communication\cite{Ben}. Recently, experimental works have
produced photons which are in the entangled states and have been
used in showing violation of Bell's inequality\cite{Asp} and
quantum teleportation\cite{Bou}. However, the generation of
entangled states of electrons have only been suggested
theoretically\cite{Los}. By realizing the fact that the long spin
dephasing time of electrons in semiconductor, the conservation of
electron number in the non-relativistical limit and the rapid
development in the research of quantum dot systems, it seems quite
promising that electronic entanglement can provide a scalable
approach to quantum computing. \newline \indent Achieving quantum
computation requires precise manipulation of the controlled
Hamiltonian. In Loss and DiVicenzo's work\cite{Los}, they proposed
to implement quantum gates by a time dependent Heisenberg spin
Hamiltonian. More recently, generation of EPR states from
two-electron mixed state has also been proposed in quantum dots
system\cite{Oli}. As all effective interactions in condensed
matter system are basically electromagnetic, it is therefore
interesting to address the entanglement problem within the
framework of Quantum Electrodynamics(QED). This approach not only
has the advantage of relativistic invariance but also takes care
of the problem of indistinguishable property of electrons. In a
recent paper, Pachos and Solano\cite{Sol} have discussed the
problem of the relativistic invariance of entanglement by using
QED. They have also claimed to obtain the Heisenberg Hamiltonian
by considering the lowest-order scattering process of two
electrons. Since magnetic interactions are of relativistic nature,
it is important to consider all the relativistic effects to the
same order. Due to the fact that electrons are fermions, the
antisymmetric nature of the state vector leads to interesting
results which do not arise for non-identical particles. Moreover,
there is also another aspect of entanglement which needs to be
addressed. Namely, the effects of interaction on the evolution of
entanglement states should be investigated. The problem of
entanglement and identical particles has also been addressed by
Omar et al\cite{Oma}, however they concentrate on transferring
entanglement from the internal to the spatial degrees of freedom
without considering interaction. In this work, we analyze the
generation and evolution of entangled electron pair within the
non-relativistic(NR) expansion by including the interaction
effects to the order of $(\frac{v}{c})^2$ and using properly
antisymmetric wave function. Consistency then requires the QED
corrections to $O(\alpha^2)$ with $\alpha$ being the fine
structure constant. Therefore, for completeness, the two-photons
exchanged processes are also discussed in this work.
\newline \indent

%\section{Review of Schmidts Decomposition and Effective Hamiltonian}

\indent \ \ \ \ Due to the non-relativistic approach of this work,
the state of the two-electron system can be expressed by a wave
function. Furthermore, with electrons being identical fermions,
the total wave function of the electron system is required to be
totally antisymmetric. In discussing scattering process the total
wave function of two electrons $\Psi$ can be expressed by the
direct product of spatial wave function $\psi$ and spin wave
function $\chi$ : $\Psi=\psi\otimes\chi$. For the moment it is
appropriate to introduce the concept of Schmidts decomposition on
the spin wave function such that the entanglement analysis can be
performed effectively.
\newline\indent
The wave function of any bipartite system is expressed as a double
sum. The Schmidts decomposition procedure asserts that the double
sum can be expressed as a single sum by local unitary
transformation. For example, the spin state vector of two spin-1/2
particles is:\begin{equation}
|\chi\rangle=\sum_{nm}C_{nm}|\zeta_n\rangle_A|\omega_m\rangle_B,
\end{equation}
where $\{|\zeta_1\rangle, |\zeta_2\rangle\}_A$ and
$\{|\omega_1\rangle, |\omega_2\rangle\}_B$ are two sets of
orthonormal basis which belong to the spin Hilbert space of the
bipartite system $A$ and $B$. By applying the Schmidts
decomposition procedure $|\chi\rangle$ can be reduced as :
\begin{equation}
|\chi\rangle=a_1|\eta_1\rangle_A|\xi_1\rangle_B+a_2
|\eta_2\rangle_A|\xi_2\rangle_B.
\end{equation}
Here, $\{|\eta_1\rangle, |\eta_2\rangle\}_A$ and $\{|\xi_1\rangle,
|\xi_2\rangle\}_B$ are the two possible incomplete orthonormal
basis of the Hilbert space, where $\{a_1, a_2\}$ are Schmidts
coefficients which satisfy the normalization condition
$\displaystyle{\sum_{i=1}^2a_i^2=1}$. This decomposition procedure
is particularly useful for expressing entangled state. For this
two-spin system, if all $a_i$  are non-vanishing
 then the system is spin entangled which means the state
vector $|\chi\rangle$ can not be expressed as a direct product
$|\eta'\rangle_A|\xi'\rangle_B$. By picking a special direction as
z-axis, one can define a conventional basis
$\{\mid\uparrow\rangle, \mid\downarrow\rangle\}$ along this axis
such that any spin state vector can be expanded as linear
combination of these basis vectors. Denoted a symmetrized and
entangled spin vector as $|\chi^E_S\rangle$, the general form
written in the conventional basis is:

\begin{equation}
|\chi^E_S\rangle=C\mid\uparrow\uparrow\rangle+
D(\mid\uparrow\downarrow\rangle+\mid\downarrow\uparrow\rangle)+
G\mid\downarrow\downarrow\rangle~~\mbox{ with }D^2\neq CG.
\end{equation}
%It is easy to see that, when $a_1=a_2$,
%$|\chi^E_S\rangle=C|\uparrow\uparrow\rangle+E|\downarrow
%\downarrow\rangle$.
By tuning the coefficients C, D and G, three of the Bell states,
$|\Phi^{\pm}\rangle=\frac{1}{\sqrt2}(\mid\uparrow\uparrow\rangle\pm
\mid\downarrow\downarrow\rangle)$ and
$|\Psi^+\rangle=\frac{1}{\sqrt2}(\mid\uparrow\downarrow\rangle+
\mid\downarrow\uparrow\rangle)$ can be obtained from Eq.(3). If
$D^2=CE$ then the state is not entangled, that is to say by local
unitary transformation the state can be written as a direct
product. On the other hand, for the antisymmetric and entangled
spin vector $|\chi^E_A\rangle$ one can easily prove that the state
is the remaining Bell state $|\Psi^-\rangle$,
\begin{equation}
|\chi^E_A\rangle=|\Psi^-\rangle=\frac{1}{\sqrt 2}
(\mid\uparrow\downarrow\rangle-\mid\downarrow\uparrow\rangle),
\end{equation}
It is also noted that if the spin degree of the system is not
entangled, then one of the Schmidts coefficients must vanish, and
the other is equal to one.
\newline\indent By properly anti-symmetrized the total wave
function and choosing the Coulomb gauge, the interaction
potential\cite{Lan} to the $v^2/c^2$ is:
\begin{equation}
U=U_C+U_{SL}+U_{SS},
\end{equation}
where $U_C$, $U_{SL}$ and $U_{SS}$, denotes the Coulomb
interaction with relativistic correction, spin-orbital
interaction, and spin-spin interaction respectively. The form of
these three interactions are as following:
\begin {eqnarray}
U_C=\frac{e^2}{r}-\frac{{\pi}e^2\hbar^2}{4c^2m^2}\delta({\bf{r}})
-\frac{e^2}{2m^2c^2r}\{{\bf{p_1}}\cdot{\bf{
p_2}}+\frac{{\bf{r}}\cdot({\bf{r}}\cdot{\bf{p_1}}){\bf{p_2}}}{r^2}\}
\\
U_{LS}=-\frac{e^2{\hbar}}{4m^2c^2r^3}\{{\bf{r}}\times({\bf{p_1}}
-{\bf{p_2}})\cdot({\bf{{\sigma_1}}}+{\bf{{\sigma_2}}})\}
\\
U_{SS}=\frac{e^2\hbar^2}{4m^2c^2}\{[\frac{8\pi}{3}\delta({\bf{r}})
+\frac{1}{r^3}]({\bf{\sigma_1}}\cdot{\bf{\sigma_2}})
-\frac{3({\bf{\sigma_1}}\cdot{\bf{r}})
({\bf{\sigma_2}}\cdot{\bf{r}})}{r^5}\}.
\end{eqnarray}
It is noted that all the correction terms are of the same order
which is $v^2/c^2$ correction to the Coulomb potential
$\frac{e^2}{r}$. By applying these interacting potentials in the
Born approximation, a systematic analysis on entanglement can be
proceeded. However, it is not consistence if one stops at the
first Born approximation of order $\alpha$. As well known in
atomic physics problem, the energy correction due to the
relativistic effect is the same order as $\alpha^2$. Furthermore,
for electrons in metal, the Fermi velocity $v_F$ is of the order
of $10^6m/s$ and hence the ratio $v_F/c$ is the same order as
$\alpha$. Therefore, consistency requires the consideration of the
second Born term which is $O(\alpha^2)$. This higher order
analysis will be discussed later.
\newline
\indent In order to address the generation and evolution of
entangled states of two electrons, the effects of the spatial wave
function must be considered. According to the principle of quantum
mechanics on identical particles, the total wave function is
either symmetric or antisymmetric depending on the nature of the
system. Therefore any sensible discussions on entanglement must
take into account the spatial property. This is in contrast to the
conventional discussion on entanglement where only the spin states
are involved. With this in mind the problem can be posed in the
following way. If the system is prepared without spin
entanglement, then after scattered by the above interaction
potentials, is it possible to generate a spin entangled state? The
more interesting question is what type of spin entanglement state
is robust during the scattering process? These questions are
addressed in the next section.
%\section{The Scattering Process}
\indent Let us consider the two electron scattering process. The
initial wave function can be prepared as the simultaneous
eigenfunction of the Hamiltonian and total momentum operators. The
three possible different forms of the wave function are:
\begin{eqnarray}
|\Psi^{(1)}\rangle=|\psi_A(\bf{p_1},
\bf{p_2})\rangle\otimes|\chi^{\not{E}}_S\rangle \nonumber\\
|\Psi^{(2)}\rangle=|\psi_A(\bf{p_1},
\bf{p_2})\rangle\otimes|\chi^E_S\rangle \\
|\Psi^{(3)}\rangle=|\psi_S(\bf{p_1},
\bf{p_2})\rangle\otimes|\chi^E_A\rangle. \nonumber
\end{eqnarray} \noindent Here, $|\psi_S({\bf{p_1}},
{\bf{p_2}})\rangle$ and $|\psi_A({\bf{p_1}}, {\bf{p_2}})\rangle$
represent respectively the symmetric and antisymmetric wave
functions of the system with momenta ${\bf{p_1}}$ and
${\bf{p_2}}$. (Such initial states in principle can be created by
doubly ionizing the two electrons in the ground state of the
helium atom.) The explicit form of these functions
$|\psi_S\rangle$ and $|\psi_A\rangle$ are :
\begin{eqnarray}
|\psi_S({\bf{p_1}},
{\bf{p_2}})\rangle=\frac{1}{(2\pi\hbar)^{3/2}}\{e^{\frac{i}{\hbar}(\bf{p_1\cdot
x_1}+p_2\cdot x_2)}+e^{\frac{i}{\hbar}(\bf{p_2\cdot x_1}+p_1\cdot
x_2)}\} \nonumber\\
|\psi_A({\bf{p_1}},
{\bf{p_2}})\rangle=\frac{1}{(2\pi\hbar)^{3/2}}\{e^{\frac{i}{\hbar}(\bf{p_1\cdot
x_1}+p_2\cdot x_2)}-e^{\frac{i}{\hbar}(\bf{p_2\cdot x_1}+p_1\cdot
x_2)}\}. \nonumber \end{eqnarray} The form of $|\chi^E_S\rangle$
and $|\chi^E_A\rangle$ are given by Eqs. (3) and (4) respectively.
The state $|\chi^{\not{E}}_S\rangle$ representing the no entangled
spin state is :
\begin{equation}
|\chi^{\not{E}}_S\rangle=C\mid\uparrow\uparrow\rangle+
D(\mid\uparrow\downarrow\rangle+\mid\downarrow\uparrow\rangle)+
G\mid\downarrow\downarrow\rangle~~\mbox{ with }D^2=CG.
\end{equation}

Let the initial and final states of the system be denoted
respectively as $|\Psi^{(k)}_i\rangle$ and $|\Psi^{(k)}_f\rangle$
where $k$ represents any one of the states of Eq.(9). Due to the
fact that some of the terms of the interaction potential $U$, such
as $U_C$, $U_{LS}$ and $\{\bf{\sigma_1}\cdot\bf{\sigma_2}\}$, are
commuting with total spin ${\bf{S}}={\bf{s_1}}+{\bf{s_2}}$,
therefore the scattering process through these terms is spin
angular momentum conserving. Even though the last term of Eq.(8)
does not commute with the total spin $\bf{S}$, however when acting
on either symmetric or antisymmetric spin state, the resulting
state retains the same symmetry property. This is due to the fact
that this term is symmetric. Explicitly the form of the resulting
states are:
\begin{equation}
 (\bf{\sigma_1}\cdot\bf{r})(\bf{\sigma_2}\cdot\bf{r})
 |\chi_A\rangle = -\frac{r^2}{4}|\chi_A\rangle \end{equation}
\begin{equation}
 (\bf{\sigma_1}\cdot\bf{r})(\bf{\sigma_2}\cdot\bf{r})
 |\chi_S\rangle = \omega_1\mid
 \uparrow\uparrow\rangle+\omega_2(\mid\uparrow\downarrow\rangle+
 \mid\downarrow\uparrow\rangle)+\omega_3\mid\downarrow\downarrow\rangle
\end{equation}

\noindent where $\omega_i$ are some spatial factors. Then it is
obvious that the scattering between symmetric and antisymmetric
spin state by this term produces zero transition amplitude. As a
result the total scattering amplitude of the initial state
$|\Psi^{(1)}_i\rangle$ is :
\begin{equation}
\langle\Psi^{(k)}_f|U|\Psi^{(1)}_i\rangle \left\{\begin{array}{ll}
=0, & {k=3} \\
\neq 0, & {otherwise.}
\end{array}\right.
\end{equation}

This result shows that as the initial state being no spin
entanglement, the probability for creating a spin singlet
entangled state $|\Psi^{(3)}_f\rangle$ is zero. This result also
holds for the initial state being $|\Psi^{(2)}_i\rangle$, which is
an entangled spin triplet state. On the other hand, by taking the
spin singlet state $|\Psi^{(3)}_i\rangle$ as initial state, one
can make use of the complex conjugate of the above results to
conclude that the final state can only be the same spin entangled
state. Furthermore, the results also imply if we prepare an
initial spin state as either $|\chi^{\not{E}}_S\rangle$ or
$|\chi^E_S\rangle$, the interaction potential $U$ can change the
degree of spin entanglement. For example, the spin triplet
entangled state $|\chi^E_S\rangle$ can scatter into a no spin
entangled state $|\chi^{\not{E}}_S\rangle$. Therefore one has an
important conclusion that, by taking QED into account, the degree
of entanglement is not an invariant concept during time evolution.
From above discussions, one may conclude that, due to the
stability of entanglement of the spin-singlet state,
implementation of spin-singlet state in quantum computation is
more practical.

%\section{High-Order Corrections}
\indent In order to justify the above conclusion, it is necessary
to establish the fact that the high-order corrections do not spoil
the result. Furthermore, there is also another important issue
which requires at least to include the next order correction. This
was mentioned earlier that all correction terms are of the same
order $O(v^2/c^2)$ correction to the Coulomb potential
$\frac{e^2}{r}$. Since the magnitude of $v^2/c^2$ is the same as
$\alpha^2$, the second Born approximation by the interaction
potential $U$ should be included in the discussion.
\newline\indent
A complete $\alpha^2$ correction to the scattering amplitude must
include also the radiative correction to the coupling constant
$\alpha$ and the electron mass. However, since this work is only
considering the non-relativistic expansion, these radiative
effects do not alter the results of scattering and will be
neglected. The second Born approximation can be obtained directly
from calculating the corresponding Feynman diagrams which are the
ladder and crossed diagrams given in figure 1. This approach is
quite elaborated for the result of this section. Fortunately,
there exists an effective way of calculating the results within
the validity of the non-relativistic expansion. It is known that
the $\alpha^2$-correction can be obtained by a second order
potential $U^{(2)}$:
\begin{equation}
U^{(2)}=V^{(2)}_L+V^{(2)}_X, \end{equation}

%\vskip 5pt
\begin{figure}[]
\begin{center}
\includegraphics[width=90mm]{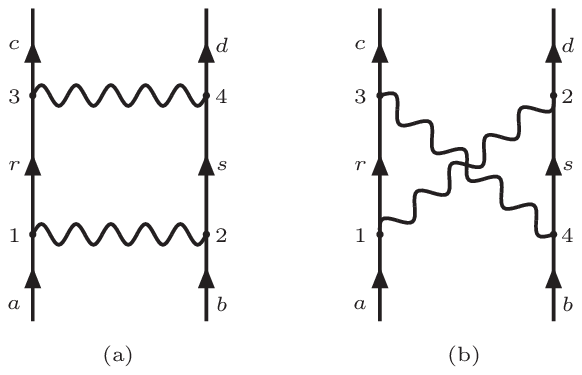}
\end{center}
\begin{center}
Figure 1. (a)The ladder and (b)the crossed diagrams of two-photon
exchange processes.
\end{center}
\end{figure}
%---------- FIGURE END ------------
%\vskip 15pt

\noindent where $V^{(2)}_L$ and $V^{(2)}_X$ are calculated from
the ladder and crossed diagrams respectively. Even though the form
of these potentials are gauge dependent, it has been shown that
the total scattering amplitudes are gauge independent\cite{Lin}.
Therefore one can chooses a convenient gauge to calculate the
amplitude. In the present study the Coulomb gauge is more
suitable, as will be shown latter, the leading relativistic
correction of the $V^{(2)}_X$ vanishes in the Coulomb gauge.

\indent To ease the discussion, the notation for the wave function
is modified slightly. All the initial states of two electrons are
written as $|ab\rangle$ and the final states as $|cd\rangle$. The
lowest order scattering amplitude $\langle cd|S^{(1)}|ab\rangle$
can be identified with a potential scattering with the interaction
potential $U$ given by Eq.(5) :
\begin{equation}
\langle cd|S^{(1)}|ab\rangle=-2 \pi
i\delta(\varepsilon_a+\varepsilon_b-\varepsilon_c
-\varepsilon_d)\langle cd|U_{12}|ab\rangle,
\end{equation}
where $\varepsilon_a$ denotes the energy eigenvalue of the free
electron of state $a$ and $U_{12}$ is the potential of vertices
$1$ and $2$ . The two-photon exchange contributions to the order
of interest, namely $O(\alpha^2)$, can be expressed as the
effective potentials which are given by :
\begin{equation}
\langle cd|V^{(2)}_L|ab\rangle=-\frac{1}{2 \pi i}\sum_{rs}\int
dz\frac{\langle
c_3d_4|U_{34}(\varepsilon_a-\varepsilon_c-z)|r_3s_4\rangle \langle
r_1s_2|U_{12}(z)|a_1b_2\rangle}
{(\varepsilon_a-\varepsilon_r-z+i\eta\varepsilon_r
)(\varepsilon_b-\varepsilon_s+z+i\eta\varepsilon_s )}
\end{equation}
and similarly for the crossed diagram
\begin{equation}
\langle cd|V^{(2)}_X|ab\rangle=-\frac{1}{2 \pi i}\sum_{rs}\int
dz\frac{\langle
c_3s_4|U_{34}(\varepsilon_a-\varepsilon_c-z)|r_3b_4\rangle \langle
r_1sd_2|U_{12}(z)|a_1s_2\rangle}
{(\varepsilon_a-\varepsilon_r-z+i\eta\varepsilon_r
)(\varepsilon_d-\varepsilon_s+z+i\eta\varepsilon_s )}.
\end{equation}
Here, $U_{ij}(cq)$ is the Fourier transform of the potential with
$q$ being the three dimensional momentum transfer. As shown in
Lindgren's work\cite{Lin}, the result of Eq.(16) can be simplified
as :
\begin{equation}
\langle cd|V^{(2)}_L|ab\rangle=\frac{1}{2}\sum_{rs}\frac{\{\langle
cd|r^{-1}_{12}|rs\rangle \langle rs|V^{BM}|ab\rangle  +\langle
cd|V^{BM}|rs\rangle \langle rs|r^{-1}_{12}|ab\rangle\}}{
(\varepsilon_a+\varepsilon_b-\varepsilon_r-\varepsilon_s)}
\end{equation}
Here, $V^{BM}$ is the generalized potential of Brown and
Mittleman\cite{Bro}
\begin{equation}
\langle rs|V^{BM}|ab\rangle=\frac{1}{2} \langle
rs|U(\varepsilon_a-\varepsilon_r)+U(\varepsilon_b-
\varepsilon_s)|ab\rangle.
\end{equation}
Since what one needs from this calculation is the order $\alpha^2$
correction, it is obvious that one should only keep the Coulomb
potential $e^2/r$ of $U$ in the above expression for $V^{BM}$.
However, due to fact that the Coulomb potential does not flip the
spin of electrons, hence $V^{(2)}_L$ does not scatter the
spin-singlet state into other spin configuration. It is now
important to show that $V^{(2)}_X$ also preserves the spin-singlet
state.
\newline\indent
For the crossed diagram, one can also simplify Eq.(17) and the
result is:
\begin{equation}
\langle cd|V^{(2)}_X|ab\rangle=-\frac{1}{2}\sum_{rs}
\frac{\{\langle cs|r^{-1}_{12}|rb\rangle \langle
rd|V^{diff}|as\rangle  +\langle cs|V^{diff}|rb\rangle \langle
rd|r^{-1}_{12}|as\rangle\}}{
(\varepsilon_a+\varepsilon_s-\varepsilon_d-\varepsilon_r)}
\end{equation}
, where $V^{diff}$ is the difference potential
\begin{equation}
\langle rs|V^{diff}|ab\rangle=-\langle
rs|U(\varepsilon_a-\varepsilon_r)-U(\varepsilon_b-
\varepsilon_s)|ab\rangle.
\end{equation}

Thus, within the same approximation, one should retain only the
Coulomb potential and therefore a vanishing $V^{diff}$ is
obtained. As a result, to the leading relativistic order,
$V^{(2)}_X$ does not contribute to the scattering process. This
completes the proof of the stability of the spin-singlet
entanglement.

%\section{Conclusions}
\indent Recently, people have tried to control electron spin as
qubit directly in some condensed matter system, such as electrons
in quantum dots. It is also claimed the interaction of such system
can be described by the Heisenberg Hamiltonian
$H=J(t){\bf{\sigma_1\cdot\sigma_2}}$. It is easy to check that the
Bell states are stable during time evolution with this
Hamiltonian. However, in this approach, spins are the only
variables of the system which, in general, can not properly
describe electrons. As pointed out earlier, it is inadequate by
neglecting the spatial wave function which is involved in the
formalism of identical particles. The effects of identical
particle on entanglement can be seen from previous discussion that
the entangled spin-triplet states, such as Bell states, can evolve
into no spin entangled state. As a result the concept of
entanglement for spin-triplet state is not practical. This is in
contrast to the case described solely by the Heisenberg
Hamiltonian. On the other hand, as shown also in this work, the
spin-singlet state which is entangled by nature is stable to a
high degree of accuracy. Therefore, for any practical
implementation of entangled state with spin, the spin-singlet
state should deserve more attention. Furthermore, the result
obtained here by considering the antisymmetric wave function,
implies that the spin-singlet entangled state can not be generated
from the other spin configuration regardless of its state of
entanglement .
\newpage
\noindent{\bf Acknowledgement.}
\newline
This work is supported by $\bf NSC$91-2112-M-006-008.
\newline
\newline

\renewcommand{\baselinestretch}{0.6}


\begin{thebibliography}{99}

%\begin {thebibliography}{99}
\bibitem{Ein} A. Einstein, N. Rosen and B. Podolsky, Phys. Rev.
$\bf{47}$, 777 (1935).
\bibitem{Ben} C. H. Bennett and G. Brassard, in \textit{Proceedings of the
IEEE Conference on Computers, Systems and Signal Processing,
bangalore, India }(IEEE, New York, 1984), p. 174; A. K. Ekert,
Phys. Rev. Lett. $\bf{67}$, 661 (1991)
\bibitem{Asp} A. Aspect, P.
Grangier, and G. Roger, Phys. Rev. Lett. $\bf{47}$, 460 (1981); G.
Weihs \textit{et al., ibid.} $\bf{81}$, 5039 (1998).
\bibitem{Bou} D. Bouwmeester \textit{et al.}, Nature (London) $\bf{390}$, 575
(1997); D. Boschi \textit{et al} Phys. Rev. Lett. $\bf{80}$ 1121
(1998); A. Furusawa \textit{et al.}, Science $\bf{282}$, 706
(1998).
\bibitem{Los} D. Loss and D. P. DiVincenzo, Phys. Rev. A
$\bf{57}$, 120 (1998).
\bibitem{Oli} W. D. Oliver, F. Yamaguchi, and Y. Yamamoto, Phys.
Rev. Lett., $\bf{88}$, 037901-1 (2002).
\bibitem{Sol} J. Pachos and E. Solano, QIC $\bf{3}$, 115, (2003), $\bf{quant-ph/0203065}$.
\bibitem{Oma} Y. Omar, N. Paunkovi\'{c}, S. Bose and V. Vedral,
Phys. Rev. A $\bf{65}$, 062305-1, (2002).
\bibitem{Sch} E. Schmidts, Math. Ann. $\bf{63}$, 433 (1907).
\bibitem{Lan} L. D. Landau and E. M. Lifshitz, \textit{Relativistic
Quantum theory part I}, P.284, (Addison-Wesley, 1971).
\bibitem{Lin} I. Lindgren, J. Phys. B $\bf{23}$, 1085(1990); C. C.
Chen, J. Phys. B $\bf{26}$, 599 (1993).
\bibitem{Bro} G. Brown, Phil. Mag. $\bf{43}$, 467 (1952); M. H.
Mittleman, Phys. Rev. A $\bf4$, 893 (1971) and Phys. Rev. A
$\bf5$, 2389 (1972).
\end{thebibliography}
\end{document}